\documentclass[acmsmall]{acmart}

\AtBeginDocument{%
  }

\setcopyright{acmlicensed}
\copyrightyear{2026}
\acmYear{2026}
\acmDOI{XXXXXXX.XXXXXXX}
\acmConference[CSCW '26]{Proceedings of the ACM on Human-Computer Interaction}{2026}{Salt Lake City, UT, USA}
\acmISBN{978-1-4503-XXXX-X/2026}

\usepackage{booktabs}
\usepackage{array}
\usepackage{caption}
\usepackage{float}

\begin{document}

\title{FaciliTrain: Practicing Facilitation Skills through AI-Simulated Group Dialogue}

\author{Hang Jiang}
\affiliation{%
  \institution{Northeastern University \& MIT}
  \city{Cambridge}
  \state{MA}
  \country{USA}}
\email{hang.jiang@northeastern.edu}

\author{Yuanxin Zhu}
\affiliation{%
  \institution{Cornell University \& MIT}
  \city{New York}
  \state{NY}
  \country{USA}}
\email{yz3477@cornell.edu}

\author{Diyi Yang}
\affiliation{%
  \institution{Stanford University}
  \city{Stanford}
  \state{CA}
  \country{USA}}
\email{diyiy@stanford.edu}

\author{Yoon Kim}
\affiliation{%
  \institution{MIT}
  \city{Cambridge}
  \state{MA}
  \country{USA}}
\email{yoonkim@mit.edu}

\author{Deb Roy}
\affiliation{%
  \institution{MIT}
  \city{Cambridge}
  \state{MA}
  \country{USA}}
\email{dkroy@mit.edu}

\author{Jad Kabbara}
\affiliation{%
  \institution{MIT}
  \city{Cambridge}
  \state{MA}
  \country{USA}}
\email{jkabbara@mit.edu}

\begin{abstract}
Skilled facilitation supports inclusive small-group dialogue, but deliberate practice is hard to scale: it depends on expert coaches, live practice partners, and iterative feedback. We present \textit{FaciliTrain}, a voice-based training system in which learners step into the facilitator role of an AI-simulated multi-participant conversation, apply five evidence-based techniques, and receive structured AI feedback to support reflection. We report findings from a mixed-methods study with 24 participants, conducted as a formative study (N = 12) and a controlled pilot (N = 12; 6 treatment, 6 control). Both conditions achieved comparable accuracy on a live evaluation task, though treatment participants' self-rated comfort declined significantly while control participants' comfort improved (p = .018). Reflexive thematic analysis identifies four themes: the taxonomy externalizes implicit facilitation intuitions; Making Connections is the most cognitively demanding technique; voice acts as a deliberate-response forcing function; and participants overwhelmingly preferred AI feedback over self-practice. We discuss design implications for voice-based, AI-supported interpersonal skill training at scale.
\end{abstract}



\maketitle

\section{Introduction}

Effective facilitation helps small-group conversations remain inclusive, constructive, and grounded in participants' lived experience. In polarized or cross-difference contexts, skilled facilitation can determine whether dialogue produces understanding or stalemate~\cite{schroeder-etal-2024-fora}. Yet facilitator training is resource-intensive: it depends on expert coaches, live practice partners, and iterative feedback cycles that are hard to scale.

Recent work has positioned large language models (LLMs) as practice partners for interpersonal skills, with Rehearsal demonstrating that LLM-based conflict resolution practice transfers to real-world outcomes~\cite{rehearsal} and a growing set of systems addressing peer counseling, motivational interviewing, and empathy training~\cite{yang2024social, mohanty-etal-2025-bridging, coempa2026, guevarra2025gloss}. Two limitations of this literature are central to facilitation. First, most existing systems target one-on-one, text-based interaction, while small-group facilitation imposes distinct demands: managing multiple voices in real time, bridging perspectives, and responding under time pressure with spoken language. Voice-based agents specifically remain underexplored as communication training partners~\cite{cava2025}. Second, a parallel line of work positions AI \textit{as} the facilitator~\cite{haqbeen2020promoting, oyama2021ai, sahab2024conversational, keeper}, supporting group process directly rather than building human facilitator capacity. Yet face-to-face and small-group dialogue, where facilitation matters most, are unlikely to be delegated to AI. Spreading human facilitation capacity creates a multiplier effect that AI-as-facilitator cannot replicate: each trained facilitator goes on to support conversations in their own community. We therefore take a different approach: using AI to \textit{train} human facilitators at scale.

We present \textbf{FaciliTrain}, a voice-based system in which users step into the facilitator role of an AI-simulated multi-participant conversation, practice five evidence-based techniques drawn from intergroup dialogue research~\cite{schroeder-etal-2024-fora}, and receive structured AI feedback to support reflection. We contribute the system design, findings from a 24-participant mixed-methods study (formative N = 12; controlled pilot N = 12), and four qualitative themes with design implications for voice-based, AI-supported facilitation training. We ask: \textbf{RQ1:} Does FaciliTrain support facilitation technique acquisition across structured practice conditions? \textbf{RQ2:} What are learners' experiences with voice-based, AI-simulated group dialogue as a training context? \textbf{RQ3:} How does AI-generated feedback shape learners' self-perception and reflection compared to self-practice alone?

\section{The FaciliTrain System}

\begin{figure}[!ht]
\centering
\includegraphics[width=0.9\linewidth]{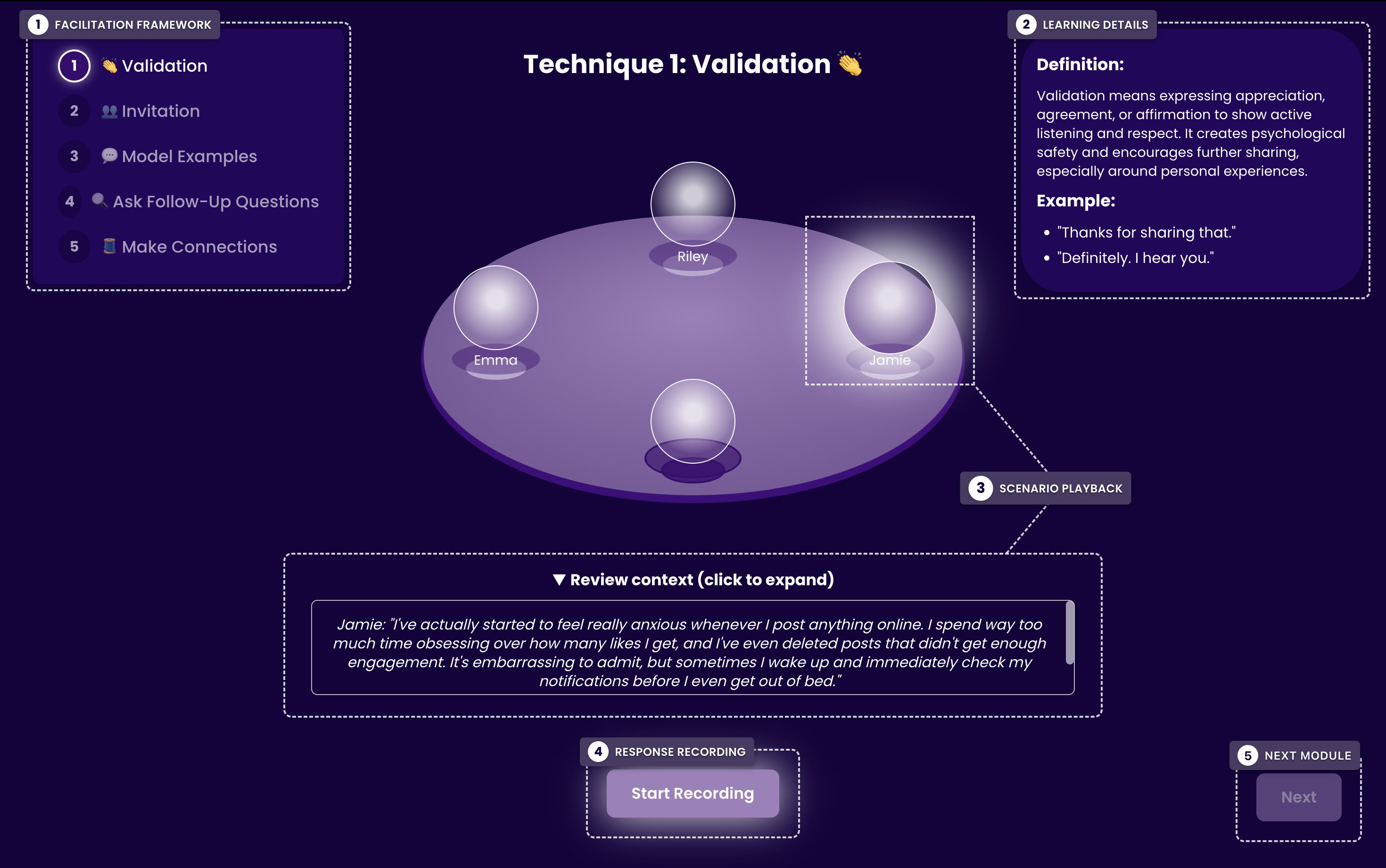}
\caption{FaciliTrain interaction flow: (1) Technique Taxonomy overview; (2) technique definition and examples; (3) scenario playback; (4) voice response with AI feedback; (5) modeled example before advancing.}
\label{fig:workflow}
\end{figure}

FaciliTrain is designed around three goals grounded in learning theory and prior work on skill acquisition and facilitation. (1)~\textit{Experiential practice}: deliberate, repeated practice is central to skill acquisition~\cite{rehearsal, yang2024social}, and facilitation in particular requires doing rather than observing~\cite{schroeder-etal-2024-fora}. (2)~\textit{Authentic facilitation demands}: voice interaction, multi-party dynamics, and unscripted response pressure are constitutive features of real facilitation, not incidental modality choices; voice-based agents are specifically underexplored as communication training partners~\cite{cava2025}. (3)~\textit{Reflection over grading}: learning from practice requires reflection on the gap between intended and observed behavior; feedback is therefore framed to support iteration rather than enforce a single correct response~\cite{yang2024social}. The system teaches five techniques from intergroup dialogue practice~\cite{schroeder-etal-2024-fora}: \textbf{\textit{Validation}}, \textbf{\textit{Invitation}}, \textbf{\textit{Modeling Examples}}, \textbf{\textit{Ask Follow-Up Questions}}, and \textbf{\textit{Making Connections}}. For each, users review a definition with examples and expert audio modeling, listen to an AI-generated group conversation among three virtual participants, respond verbally as the facilitator, and receive AI feedback that names the techniques present, explains why, and offers a reflection prompt (Figure~\ref{fig:workflow}). Users may retry up to three times before a modeled expert response is played. After all five modules, participants apply techniques to a novel evaluation scenario without scaffolding.

\begin{figure}[!ht]
\centering
\includegraphics[width=1.0\linewidth]{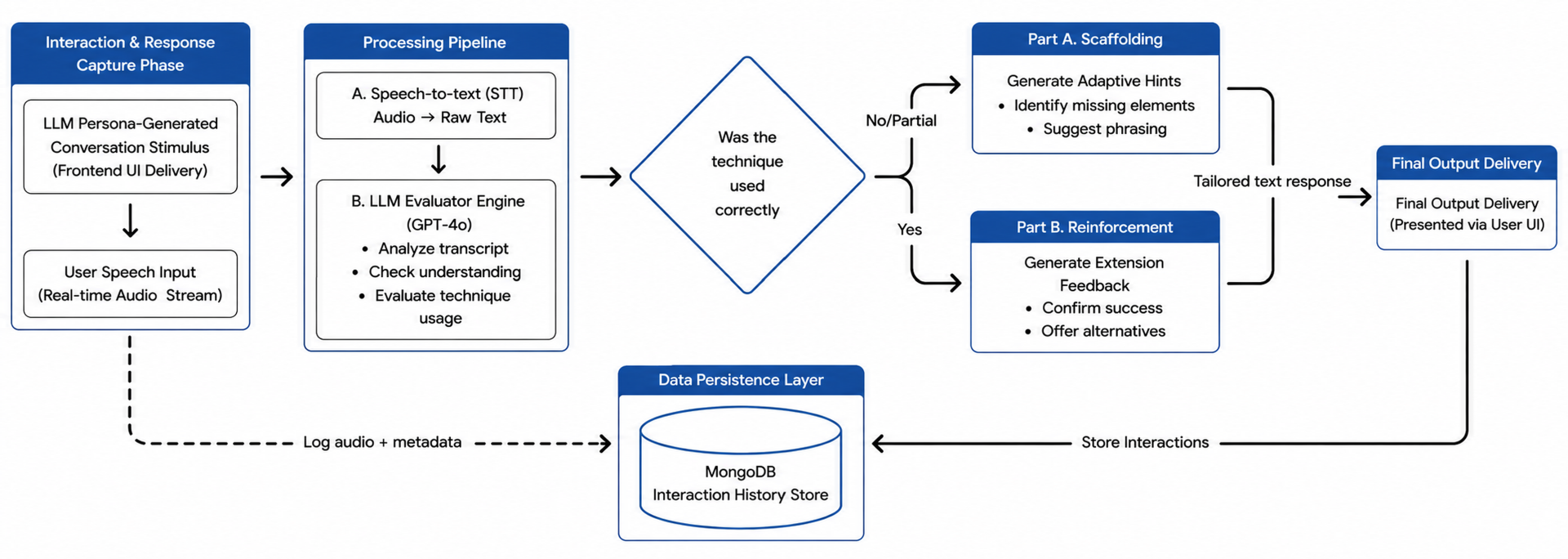}
\caption{FaciliTrain system architecture.}
\label{fig:architecture}
\end{figure}

The system architecture integrates three components (Figure~\ref{fig:architecture}): a voice interface that presents scenarios via OpenAI \texttt{tts-1} and captures speech; a speech understanding module that transcribes with \texttt{whisper-1} and classifies technique use with GPT-4o; and a feedback generation module that produces technique-level, reflection-oriented responses rather than binary correctness judgments.

\section{User Study}

We conducted a mixed-methods, IRB-approved study with 24 MIT community members in two stages.

\subsection{Formative Study (N = 12)}

The formative study (all 12 participants in the AI-feedback condition) focused on usability, yielding two insights that drove system refinements. First, participants found the single-answer multiple-choice evaluation artificially restrictive (\textit{``there is no single right way to facilitate''}), so we redesigned the evaluation to allow selecting one or more techniques per scenario. Second, participants who passed a technique on their first attempt felt over-constrained by mandatory secondary practice (\textit{``I thought my answer was good enough. Why did the feedback tell me it needs to be improved further?''}), so we made secondary practice optional after initial success.

\subsection{Controlled Pilot (N = 12)}

Participants were randomly assigned to \textit{treatment} (AI feedback, n = 6; P1 to P6) or \textit{control} (self-practice only, n = 6; P7 to P12). Groups differed in self-reported experience (treatment: 2 no experience, 2 beginner, 2 intermediate; control: 3 no experience, 3 beginner). Sessions ran over Zoom and lasted approximately 60 minutes; participants received a \$40 gift card.

Each session followed four phases: a pre-study survey (5 min); training through all five techniques (20 min), with treatment participants receiving AI feedback after each verbal response and control participants self-reflecting without feedback; a 15-minute evaluation task with three AI personas on an immigration dialogue topic; and a post-study survey and interview in which participants were shown the alternative condition.

\subsection{Data and Analysis}

We collected pre- and post-session surveys, session transcripts (training think-aloud and a 10-minute post-study interview), and facilitator responses from the evaluation task.

\textbf{Quantitative analysis.} Two researchers independently annotated all 60 facilitator responses for technique use (inter-rater F1 = 0.93, SD = 0.13); disagreements were resolved by consensus to form a gold standard. Each participant's responses were scored against this standard using F1. Between-condition comparisons of pre-to-post change scores and evaluation accuracy used two-tailed Mann-Whitney U tests.

\textbf{Qualitative analysis.} We followed reflexive thematic analysis~\cite{braun2006using}: two researchers independently coded six transcripts, developed a codebook, applied it to all 12 participants, and finalized themes through iterative discussion.

\section{Findings}

Addressing RQ1, both conditions achieved comparable technique accuracy on the live evaluation task (Treatment F1 = 0.63, SD = 0.12; Control F1 = 0.64, SD = 0.04; p = .75), and neither condition showed significant pre-to-post self-rated skill gains. We find one significant quantitative result: treatment participants' self-rated comfort declined after training ($\Delta$ = $-$0.67), while control participants' comfort improved ($\Delta$ = +0.33; p = .018, rank-biserial r = .78); notably, groups differed at baseline (treatment M = 4.33 vs.\ control M = 3.00, p = .023), complicating causal attribution. Given the small sample (n = 6 per condition), we interpret null results as inconclusive rather than evidence of no difference, and focus on four qualitative themes.

\subsection{The Technique Taxonomy Externalizes Implicit Facilitation Intuitions}

Addressing RQ2, all 12 controlled-pilot participants described already performing facilitation behaviors intuitively, validating speakers, inviting quieter voices, connecting themes, yet without naming or consciously deploying them. The system's in-context labeling made the taxonomy visible during live practice, surfacing links between action and concept that participants had not previously articulated. \textit{``People use them very frequently, but I didn't know they had... names and explanations and examples''} (P1, Treatment). The taxonomy's value is making these implicit behaviors visible, not introducing new ones. We tentatively conjecture that this naming effect may support transfer by making existing behaviors more deliberate and available for reflection.

\subsection{Making Connections Is the Most Demanding Technique}

Every participant identified Making Connections as the most cognitively taxing technique (RQ2). Unlike the others, which involve responding to a single speaker, Making Connections requires holding multiple prior contributions in working memory and constructing a meaningful relational bridge across them. P5 (Treatment) captured the difficulty: \textit{``that's the hardest thing to do, like, bridge together everybody's different viewpoints and show similarities and differences, while still validating people's views and not discounting certain people''}.

AI training logs reveal a related pattern: among treatment participants, Making Connections (2 of 6 did not pass within the allowed feedback attempts) and Modeling Examples (3 of 6) showed the highest non-completion rates. Participants could see themselves failing the AI evaluator on this technique across multiple retries. Several reported needing to re-read the full transcript before each attempt, a behavior observed in no other module. This convergence of subjective report and AI-logged behavioral data points to a concrete design opportunity: differentiated scaffolding for this technique, such as an explicit recall step or a guided compare-and-contrast prompt. We cautiously speculate that failing visibly may itself be productive, surfacing the gap between conceptual understanding and in-the-moment execution in ways self-reflection alone might not. Systems that make failure legible and recoverable could be particularly valuable for high working-memory techniques.

\subsection{Voice as a Deliberate-Response Forcing Function}

Addressing RQ2, voice interaction is valued not for authenticity alone but because it scaffolds deliberate response in a way text cannot. P9 (Control) articulated the mechanism most directly: \textit{``I usually overthink a lot about what I need to say. And so this helped me take a pause and think a lot more beforehand, versus typing is like I just start writing, and then I'm like, oh, I know I can edit this, or make it better''}. Voice removes the option of revising before submission, imposing the same in-the-moment commitment that real facilitation requires. The cognitive load of real-time response also surfaced unexpected difficulty for confident participants, with P4 (Treatment) reflecting: \textit{``I thought the facilitation was very easy. Now I understand''}. In total, 10 of 12 participants endorsed voice as the appropriate modality for facilitation training, despite initial discomfort. We conjecture that removing the safety net of editing may induce productive difficulty that accelerates toward more fluent, in-the-moment facilitation. Whether this holds across different language backgrounds is an open question.

\subsection{Participants Overwhelmingly Prefer AI Feedback to Self-Practice}

Addressing RQ3, when shown the alternative condition during the post-study interview, participants overwhelmingly preferred AI feedback to self-practice, including control participants who had practiced without it. Treatment participants, who had experienced AI feedback firsthand, were unwilling to trade it for self-practice even when critical of its accuracy. P6 (Treatment) was emphatic: \textit{``I need feedback, otherwise what is the meaning of practice?''}

Control participants were equally clear when offered the choice. P12 (Control): \textit{``feedback is really important. Reflection is nice but not enough. Feedback is what you need to improve upon yourself''}. The single exception was P7 (Control), who preferred human feedback but still ranked AI above self-practice: \textit{``over nothing, I would definitely prefer the AI feedback''}. This preference complements the null quantitative comparison: although F1 was comparable, participants treat AI feedback as a defining feature of practice rather than an interchangeable supplement. We tentatively suggest that feedback may operate less on technique accuracy and more on the experience of practice itself, providing a reference point that makes self-assessment possible and sustains motivation to engage seriously with difficult techniques.

\section{Discussion}

\textbf{Structured practice as a shared mechanism.} Both conditions achieved comparable technique accuracy (F1 $\approx$ 0.63 to 0.64), with no significant pre-to-post skill gains in either. Notably, the treatment group's higher average experience (see Limitations) makes this comparable outcome more striking. At n = 6, however, these results are preliminary: the study is underpowered, and null results should not be read as evidence of equivalence. One tentative reading is that structured practice itself may be the primary active ingredient, with AI feedback showing a non-significant trend toward higher self-rated practice value (p = .08) that may not yet manifest as measurable accuracy differences.

\textbf{Feedback, metacognition, and the comfort trade-off.} The observed comfort divergence between conditions reflects two interacting factors: a pre-existing baseline difference in self-rated comfort (treatment M = 4.33 vs.\ control 3.00, p = .023) and the experience of AI feedback making performance gaps visible in real time. For treatment participants, this visibility appeared to produce uncomfortable self-scrutiny, temporarily reducing confidence even as performance remained stable. Control participants, without a comparative reference, tended to express uncertainty differently (\textit{``I think I'm doing it, but maybe I'm not''}, P10). This pattern suggests that increased metacognitive demand is not inherently problematic, but designers should anticipate it and scaffold the emotional alongside the informational dimensions of feedback.

\textbf{Toward differentiated scaffolding.} The convergence of subjective report and behavioral failure rates on Making Connections and Modeling Examples points to two specific areas where current scaffolding appears insufficient. Both techniques impose higher working memory demands than the others, and both show elevated non-completion rates. Progressive scaffolding strategies, such as guided recall steps, explicit comparative prompts, and lower-stakes warm-up tasks, could address these demands without requiring a full system redesign.

\textbf{Simulation authenticity as a design dimension.} Eight of 12 participants noted that the AI dialogue felt stylistically uniform (\textit{``this is not how real people talk. This is how ChatGPT talks''}, P7), with unnaturally clean turn structure, absent crosstalk, and limited conflict dynamics. Notably, the same participants valued the simulation's \textit{structure} even while critiquing its realism, suggesting that authenticity and pedagogical scaffolding may be independently tunable rather than in direct tension. Real deliberative settings involve unequal airtime, pre-existing relationships, and selective information-sharing that current scripted scenarios do not yet capture. FaciliTrain represents an early step in this design space; future work should explore asymmetric engagement patterns, participant history, and conflict dynamics.

\textbf{Limitations and scope.} The study's primary constraint is statistical power (n = 6 per condition), limiting quantitative conclusions. The groups also differed in prior experience: the treatment group included two intermediate participants while the control group had none, introducing a potential confound that larger, fully balanced samples would need to address. Because both conditions experienced the same AI-simulated dialogue, this design isolates the effect of feedback but cannot speak to the value of AI simulation itself relative to other practice contexts. The MIT-affiliated sample limits generalizability beyond similar populations, and the absence of longitudinal follow-up leaves skill retention and real-world transfer open for future work.

\section{Conclusion}

We presented FaciliTrain, a voice-based AI system for practicing facilitation through multi-participant simulation. The controlled pilot finds comparable technique acquisition across AI-feedback and self-practice conditions. The qualitative findings identify four themes with design implications: taxonomies externalize implicit facilitation intuitions; Making Connections requires differentiated scaffolding; voice scaffolds deliberate response; and learners treat AI feedback as a defining feature of practice. These findings position FaciliTrain as a proof-of-concept for AI training partners that can scale human facilitation capacity.

\newpage
\bibliographystyle{ACM-Reference-Format}
\bibliography{sample-base}

\end{document}